\documentclass[a4paper]{article}

\usepackage{INTERSPEECH2021}

\usepackage[utf8]{inputenc}
\usepackage[dvipsnames]{xcolor}

\title{Layer Pruning on Demand with Intermediate CTC}
\name{Jaesong Lee$^1$, Jingu Kang$^1$, Shinji Watanabe$^2$}
\address{$^1$Naver Corporation\\$^2$Carnegie Mellon University}
\email{jaesong.lee@navercorp.com, kang.jingu@navercorp.com, shinjiw@ieee.org}

\begin{document}

\maketitle
\begin{abstract}
Deploying an end-to-end automatic speech recognition (ASR) model on mobile/embedded devices is a challenging task,
since the device computational power and energy consumption requirements are dynamically changed in practice.
To overcome the issue,
we present a training and pruning method for ASR
based on the connectionist temporal classification (CTC)
which allows reduction of model depth at run-time without any extra fine-tuning.
To achieve the goal, we adopt two regularization methods, intermediate CTC and stochastic depth,
to train a model whose performance does not degrade much after pruning.
We present an in-depth analysis of layer behaviors using singular vector canonical correlation analysis (SVCCA),
and efficient strategies for finding layers which are safe to prune.
Using the proposed method, we show that a Transformer-CTC model can be pruned in various depth on demand, improving real-time factor from 0.005 to 0.002 on GPU, while
each pruned sub-model maintains the accuracy of individually trained model of the same depth.
\end{abstract}
\noindent\textbf{Index Terms}: end-to-end automatic speech recognition, connectionist temporal classification, pruning, non-autoregressive

\section{Introduction}
\label{sec:introduction}

End-to-end automatic speech recognition (ASR) has recently become a popular approach.
It gives strong performance and simple model design compared to traditional systems like hidden Markov model.
Also, with the advances of hardware~\cite{ignatov2019ai} and software~\cite{lee2019ondevice},
it became possible to deploy the end-to-end ASR system on the mobile devices like smartphones and embedded devices.
However, developing and deploying an end-to-end ASR model on such devices is still challenging,
because of the varying computational power of the devices~\cite{ignatov2019ai}.
Recent high-end devices have very fast CPU and large memory and
some of them also have dedicated hardware like neural processor unit (NPU).
On the other hand, low-end and old-generation devices with limited computational power are also widely used.
This gives a critical problem for on-device ASR design.
A large model gives high recognition accuracy but requires significant computational cost.
Without the dedicated processor, the device consumes significant energy and time, affecting battery life~\cite{ignatov2019ai},
thus small and fast models should be needed for such devices.
This requires designing several models with different size, complicating the development and deployment of on-device ASR.

Recently, the idea of \emph{on-demand pruning} has been popularized~\cite{yu2018slimmable, yu2019universally, fan2020reducing, vu2020anywidth}.
Conventional pruning methods require three steps: 1) training large model, 2) pruning the model, and 3) fine-tuning the resulting small model.
On the other hand, pruning on demand requires \emph{no fine-tuning} after pruning.
This property is especially helpful for on-device ASR design.
Only one model is required for development and deployment,
and each device may adjust the level of computation given its own limit.
In this end, we consider the problem of on-demand pruning for ASR modeling.

Connectionist temporal classification (CTC)~\cite{graves2006connectionist} has been a widely used method for end-to-end ASR modeling~\cite{audhkhasi2017building,Audhkhasi2019,Pratap2020,quartznet,wav2vec2}, and it is especially an attractive method for the on-device ASR\@.
For low-end devices, CTC is suitable for lightweight modeling.
Compared to attentional encoder-decoder architectures~\cite{chorowski2015attention,7472621,8462506,Karita2019}
or RNN-Transducer~\cite{graves2012sequence},
CTC-based models do not require separate decoder networks, thus they require less computational cost (time and memory) both for training and inference.
Also, for high-end devices equipped with dedicated processors,
CTC with modern architectures (e.g., Transformer~\cite{NIPS2017_7181,pham2019very} and Conformer~\cite{gulati2020conformer,guo2020recent}) allows fast parallel inference,
coming from the \emph{non-autoregressive} property of CTC\@.
Also, while CTC has been regarded weaker than encoder-decoder,
various ways to improve CTC, including pre-training~\cite{wav2vec2} and regularization~\cite{lee2021intermediate,talnikar2021joint}, have been developed.
Also, there are active researches on CTC variants~\cite{wav2vec2} and on non-autoregressive modeling based on CTC~\cite{higuchi2020mask, chan2020imputer, chi2020align},
suggesting that effective pruning for CTC can also be applied to other variants or non-autoregressive models.

In this work, we consider the on-demand layer pruning problem:
training a deep neural network and removing some of the layers,
thus reducing the depth of the network, without any fine-tuning.
To achieve this, we employ two methods, intermediate CTC~\cite{lee2021intermediate} and stochastic depth~\cite{huang2016deep,pham2019,fan2020reducing}.
Although they are originally introduced as regularizers during training,
we show they can be applied to layer pruning method as well.
Note that we also explore in-place distillation~\cite{yu2019universally},
an application of knowledge distillation~\cite{li2014learning,hinton2015distilling,8461995,9003776,takahumi2020self} to on-demand pruning, but find degraded result in our ablation  study.

Our main contributions are:
\begin{itemize}
\setlength{\itemsep}{0pt}\setlength{\parskip}{0pt}\setlength{\parsep}{0pt}
\item We employ singular vector canonical correlation analysis (SVCCA) to analyze intermediate CTC and stochastic depth in context of layer pruning.
\item We show that intermediate CTC, along with stochastic depth, not only improves training of CTC models, but also provide a natural way to prune layers.
\item We develop an iterative search method augmented with intermediate CTC for pruning layers.
\item We present experimental results that the proposed model and its induced sub-models match the accuracy of individually trained models with same depth, without any fine-tuning, thus proving the effectiveness of our method.
\end{itemize}

\section{Architecture}
\label{sec:architecture}

\subsection{Connectionist Temporal Classification}

Connectionist temporal classification (CTC) solves the sequence prediction problem by introducing a monotonic alignment.
For the encoder output $x \in \mathbb{R}^{T \times D}$ of length $T$ and feature dimension $D$,
the CTC layer computes the likelihood of target sequence $y$:
\begin{equation}
P(y|x) := \sum_{a \in \beta^{-1}(y)} P(a|x),
\end{equation}
where $\beta^{-1}(y)$ is the set of possible alignments that are compatible to $y$ and are of length $T$, and $a$ is an alignment in the set.
The probability of an alignment is modeled as a factorized distribution:
\begin{align}
P(a|x) &:= \prod_t P(a[t]|x[t]) \\
\label{eq:ctc_factorized}
P(a[t] | x[t]) &:= \text{Softmax}(\text{Linear}(x[t]))_{a[t]}
\end{align}
where $a[t]$ and $x[t]$ mean the $t$-th value of $a$ and $x$ respectively.

During inference, the most probable alignment is found by greedy decoding,
allowing fast and non-autoregressive inference.



\subsection{Transformer encoder}

Transformer~\cite{NIPS2017_7181} is a multi-layer architecture with self-attention and residual connection~\cite{he2016}.
Transformer layer uses self-attention to learn global information and residual connection to help training of a deep neural network.

We use the encoder part of the Transformer architecture.
Let $L$ be the number of layers in the encoder.
The $l$-th layer computes the representation $x_l$ for given input $x_{l-1}$ by:
\begin{align}
\label{equation:transformer}
x^\text{MHA}_l &= \text{SelfAttention}(x_{l-1}) + x_{l-1} \\
\label{equation:transformer2}
x_l &= \text{FeedForward}(x^\text{MHA}_l) + x^\text{MHA}_l,
\end{align}
where $x_0$ indicates the input of the Transformer encoder.

The final representation $x_L$ is then fed to CTC layer to minimize the following loss:
\begin{equation}
\label{eq:loss_ctc}
\mathcal{L}_\mathsf{CTC} := - \log P_\mathsf{CTC}(y|x_L).
\end{equation}

\subsection{Stochastic depth and LayerDrop}

Stochastic depth~\cite{huang2016deep,pham2019} is a regularization method designed for deep residual networks~\cite{he2016}.
During training, each layer is randomly skipped or not with a given probability $p$.
For each iteration, $u$ is sampled from a Bernoulli distribution so that
the probability of $u = 1$ is $p$ and the probability of $u = 0$ is $1 - p$.
The output is computed by modifying Eqs.~\eqref{equation:transformer} and \eqref{equation:transformer2} as:
\begin{align}
\label{equation:transformer_stochdepth}
x^\text{MHA}_l &= \frac{u}{p} \cdot \text{SelfAttention}(x_{l-1}) + x_{l-1} \\
x_l &= \frac{u}{p} \cdot \text{FeedForward}(x^\text{MHA}_l) + x^\text{MHA}_l.
\end{align}
If $u = 0$, the residual parts are skipped (i.e., $x_l = x_{l - 1}$).


LayerDrop~\cite{fan2020reducing} is an application of stochastic depth to layer pruning.
After training the model with stochastic depth, removing some layers from the model gives new smaller sub-model which also has reasonable performance without any fine-tuning.


\subsection{Intermediate CTC}

Intermediate CTC~\cite{lee2021intermediate} is an auxiliary loss designed for CTC modeling.
It regularizes the model using an additional CTC loss attached at the intermediate layer of the encoder.

Let $l_1, \ldots, l_K$ be the $K$ positions ($K < L$) of the intermediate layers.
The intermediate loss is defined as:
\begin{equation}
\label{eq:loss_inter}
\mathcal{L}_{\mathsf{InterCTC}} := \frac{1}{K} \sum_k - \log P_\mathsf{CTC}(y | x_{l_k}).
\end{equation}
Then, the training objective is defined by combining Eqs.~\eqref{eq:loss_ctc} and \eqref{eq:loss_inter}:
\begin{equation}
\mathcal{L} := (1 - w) \mathcal{L}_{\mathsf{CTC}} + w \mathcal{L}_{\mathsf{InterCTC}}
\end{equation}
with a hyper-parameter $w$.

Intermediate CTC has very small overhead during training, because the intermediate representation $x_{l_k}$ is naturally obtained when the final output $x_L$ is computed.
The only additional cost is to compute CTC loss for the given representation, which is much smaller than the cost of the encoder.
\cite{lee2021intermediate} explores various choices for the intermediate layer positions,
and concludes that it is sufficient to use one layer ($K = 1$) in the middle ($l_1 = \lfloor L / 2 \rfloor$) for the regularization purpose.
We revisit the effect of variants at Section~\ref{sec:layer_similarity_analysis}.

Note that CTC and intermediate CTC share the same linear projection layer of Eq.~\eqref{eq:ctc_factorized},
as intermediate CTC is treated as a regular CTC loss except that all of the encoder layers after the intermediate layer are skipped.
This property is used at Section~\ref{sec:intermediate_ctc_layer_pruning}.

\section{Layer similarity analysis}
\label{sec:layer_similarity_analysis}

Intuitively, if a layer of the model can be pruned without much degradation of accuracy,
it would mean the next layer behaves similarly regardless that the layer is removed or not.
Therefore, measuring ``similarity'' of layer behavior would give a good intuition for layer pruning.
We measure the layer similarity using singular vector canonical correlation analysis (SVCCA)~\cite{raghu2017svcca}.

SVCCA accepts two matrices and finds two linear projections which maximize correlation between the projected matrices.
Then it averages the correlation coefficients into a scalar value.
The result, mean SVCCA similarity, indicates the similarity of two matrices up to linear transformation.
The similarity of two layers can be measured by computing the mean SVCCA similarity of the layer outputs.

We first collect outputs of all layers from each input of the validation set.
For each layer, the outputs are concatenated into a single matrix.
Then, for each pair of layers, we compute the mean SVCCA similarity of the corresponding matrices.
We also collect the inputs of the first layer and compute the similarity between it and the other layers as well.
The input is denoted as ``0th layer'' in the following figures.


\subsection{Effect of intermediate CTC and stochastic depth}

\begin{figure}[t]
  \centering
  \includegraphics[width=0.8\linewidth]{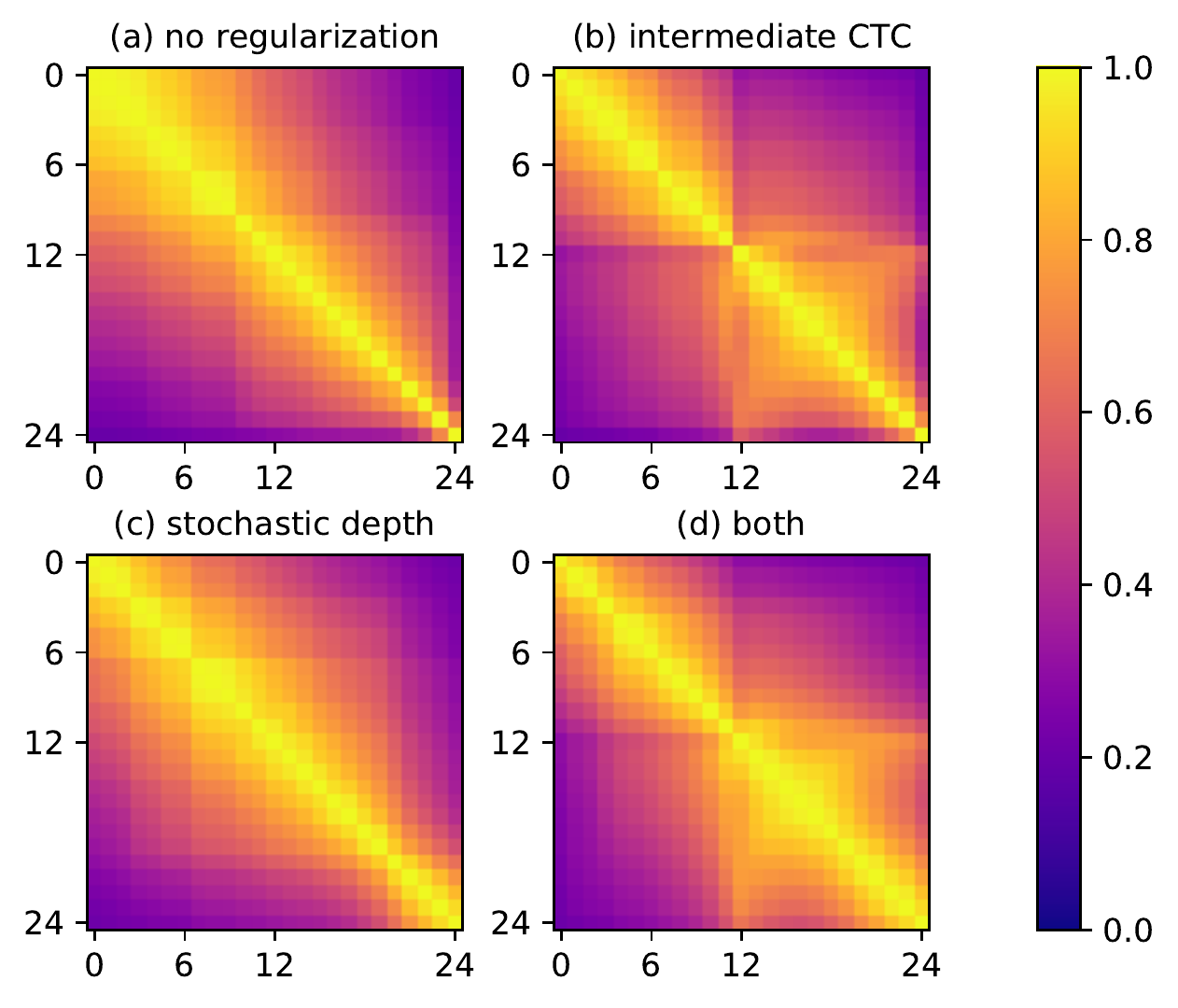}
  \caption{SVCCA similarity matrices for various regularizations.
  See Section~\ref{sec:svcca_regularizations} for details.}
  \label{fig:svcca_fig1}
  \vspace{-4mm}
\end{figure}

We first investigate how the two regularization techniques, stochastic depth and intermediate CTC, affect layer similarity.
We train a 24-layer Transformer CTC model with four different configurations: (a) baseline, (b) with intermediate CTC ($w = 0.3$, at 12th layer), (c) with stochastic depth, and (d) with both.
Figure~\ref{fig:svcca_fig1} shows the similarity matrices of the four models.

\label{sec:svcca_regularizations}

From Figures~\ref{fig:svcca_fig1} (a) and (b),
we observe that intermediate CTC increases similarity of layers between the intermediate one (12th) and the final one (24th),
at the cost of decreasing similarity of the first layers a bit.
Especially, the intermediate layer is significantly similar to last layer and its nearby layers.

On the other hand, Figure~\ref{fig:svcca_fig1} (c) shows that stochastic depth evenly increases the similarity of the last half of layers.
Especially we see the neighboring layers have very high similarities.
This can be explained by that the layer should behave similarly regardless of whether its previous layer has been skipped or not.

Finally, Figure~\ref{fig:svcca_fig1} (d) shows the combination of two regularizations greatly increases similarity among all layers between the intermediate layer and the last layer.
This gives us an insight that the both regularizations would help pruning the layers.

\subsection{Effect of intermediate CTC variants}
\label{sec:svcca_variants}
\begin{figure}[t]
  \centering
  \includegraphics[width=0.7\linewidth]{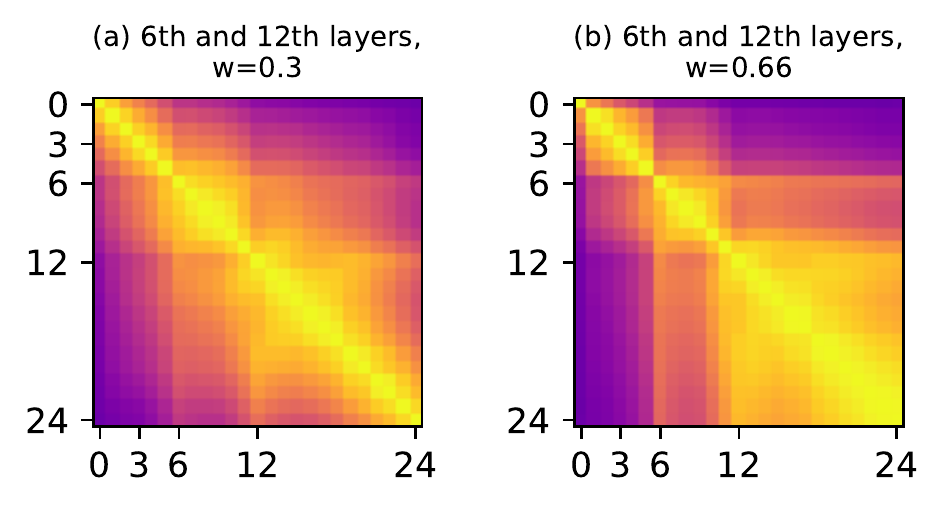}
  \caption{SVCCA similarity matrices for intermediate CTC variants.
  See Section~\ref{sec:svcca_variants} for details.}
  \label{fig:svcca_fig2}
  \vspace{-4mm}
\end{figure}
The position variant of intermediate CTC is first explored by~\cite{lee2021intermediate},
concluding that its impact to the accuracy is minimal.
However, we find the variant gives big difference on layer similarity and layer pruning.

First, we put an additional intermediate layer at the 6th layer, in addition to the 12th one.
As shown at Figure~\ref{fig:svcca_fig2} (a), it increases the similarity of layers below the 12th layer, compared to Figure~\ref{fig:svcca_fig1} (d).
This suggests that multiple branch variant affects the performance of pruning.

We then investigate the effect of the weight for the intermediate loss.
\cite{lee2021intermediate} used $w = 0.3$ as it is sufficient to enhance the model performance,
but it may not be an optimal value for pruning, especially if there are multiple intermediate layers.
Thus, we give equal weight to the two intermediate layers and the final layer, i.e., $w = 0.66 \approx 2/3$.
Figure~\ref{fig:svcca_fig2} (b) shows that the increase of intermediate weight helps overall similarity.

From the observations, we use the newly found hyper-parameters for the main experiments at Section~\ref{sec:experiments}.

\section{Layer pruning on demand}
\label{sec:layer_pruning}

Based on the layer similarity analysis of Section~\ref{sec:layer_similarity_analysis},
we propose two strategies to remove the encoder layers without much degradation of the overall accuracy.
First, we show that intermediate CTC, originally proposed for regularization, can be used for layer pruning as well.
Second, we present a more fine-grained pruning strategies which combines iterative search and intermediate CTC\@.
We emphasize that the strategies are for on-demand pruning, and there is no fine-tuning step after pruning.

\subsection{Intermediate CTC as layer pruning}
\label{sec:intermediate_ctc_layer_pruning}

The intermediate CTC loss is originally proposed for regularization,
and it is not used during inference.
However, in the context of layer pruning,
the intermediate layer suggests a natural strategy to induce a smaller sub-model from the original model:
remove all of the layers after the intermediate layer.

Moreover, we note that the method can be even extended to \emph{any depth},
even if the corresponding layer is not explicitly trained with the intermediate CTC loss.
It is because the linear projection layer of Eq.~\eqref{eq:ctc_factorized} is independent of specific layer and is shared by the intermediate and last layers,
enabling that it can be used by other layers.
Also, with the improved regularization presented at Section~\ref{sec:svcca_variants},
we expect layers between the intermediate layer and the last layer give reasonable performance.

For the layers used as intermediate branches during training,
we find the performance of their sub-models is as same as the one of the individually trained models of same depth.
For the other layers,
the performance is only slightly worse than one of individual models.
We show the experimental result at Section~\ref{sec:experiments}.

\subsection{Iterative layer pruning}
\label{sec:iterative_layer_pruning}

LayerDrop~\cite{fan2020reducing} presents pruning strategies for stochastic depth regularization.
The authors tried two strategies, removing every other layer and exhaustive search,
and concluded the two strategies do not give significant difference.
They also found removing consecutive layers causes significant degradation.

However, we find removing the last half gives nearly optimal performance, due to intermediate CTC\@.
Especially, the intermediate sub-model is significantly better than removing every other layer.
This suggests careful pruning strategy would be beneficial.

In this end, we suggest an iterative search augmented with the intermediate model.
We start with the original model, and iteratively decrease model depth by one.

For brevity, we denote sub-models as the set of number indicating the layer of the original model.
For example, if the 2-layer sub-model uses the 2nd and 4th layers of the original model,
we denote the model as $\{2, 4\}$.

For a given sub-model of depth $k$, we find a new sub-model of depth $k-1$ by the following steps:
\begin{enumerate}
\setlength{\itemsep}{0pt}\setlength{\parskip}{0pt}\setlength{\parsep}{0pt}
\item For each layer of the current model,
we induce a new sub-model by removing the layer from the model.
For example, from a given model $\{2, 3, 4\}$,
we induce three sub-models: $\{2, 3\}, \{2, 4\}, \{3, 4\}$.
\item We also induce an intermediate sub-model $\{1, \dots, k-1\}$ from the original model.
In the current example, the intermediate sub-model is $\{1, 2\}$.
\item Evaluate the induced sub-models using the validation set and choose the sub-model with the best accuracy.
\end{enumerate}
Step 2 ensures that intermediate sub-models are always included in search space.
Without Step 2, we found the iterative search may not discover them and sometimes perform worse.

Note that when a model is evaluated,
its intermediate sub-models can be evaluated as well with very small additional cost.
For example, evaluation of $\{2, 3, 4\}$ can be done along with evaluations of $\{2, 3\}$ and $\{2\}$.
This reduces computational cost for the search.

\section{Experiments}
\label{sec:experiments}

We use three ASR corpora for the experiments:
Wall Street Journal (WSJ)~\cite{paul1992design} (English; 81 hours),
AISHELL-1~\cite{bu2017aishell} (Chinese; 170 hours),
and TED-LIUM2~\cite{rousseau-etal-2014-enhancing} (English; 207 hours).

We use ESPnet~\cite{watanabe2018espnet} for the experiments.
For the input, we use 80-dimensional log-mel features with 3-dimensional pitch features and
apply SpecAugment~\cite{park2019specaugment} during training.
We put two convolution layers of stride 2 before Transformer layers.
For the output, we use character-based tokenization for WSJ and AISHELL-1 and
2000 subwords using SentencePiece~\cite{kudo-richardson-2018-sentencepiece} for TED-LIUM2.


\subsection{Main results}
\label{sec:main_results}

For layer pruning, we train a 24-layer Transformer model with intermediate CTC and stochastic depth.
Following the findings of Section~\ref{sec:layer_similarity_analysis},
we put intermediate branches at the 6th and 12th layers, and set weight $w = 0.66 \approx 2/3$.
The model is called as the pruning-aware model.
After the training, we induce smaller sub-models from the pruning-aware model by removing the layers up to the half,
either using an intermediate strategy or iterative strategy in Section~\ref{sec:layer_pruning}.
We do not fine-tune any sub-models after pruning.

To measure the performance of sub-models,
we train individual baseline models with the same depth from scratch.
For each baseline, we prepare two configurations:
(A) not using intermediate CTC or stochastic depth,
and (B) using both of intermediate CTC and stochastic depth following~\cite{lee2021intermediate}.
The baseline~B is more challenging for sub-models to reach,
as the regularizations are not only useful for pruning but also for improving individual models~\cite{lee2021intermediate,huang2016deep,pham2019,fan2020reducing}.

All models are trained for 100 epochs for WSJ, and for 50 epochs for AISHELL-1 and TED-LIUM2.
We emphasize that the pruning-aware model uses exactly same number of epochs as the individual baseline models do.
Thus, it only requires computational cost for training that the 24-layer baseline model does.
During testing, we use greedy decoding and do not use any external language models (LMs),
enabling fast, parallel and non-autoregressive generation.

\begin{figure}[t]
  \centering
  \includegraphics[width=\linewidth]{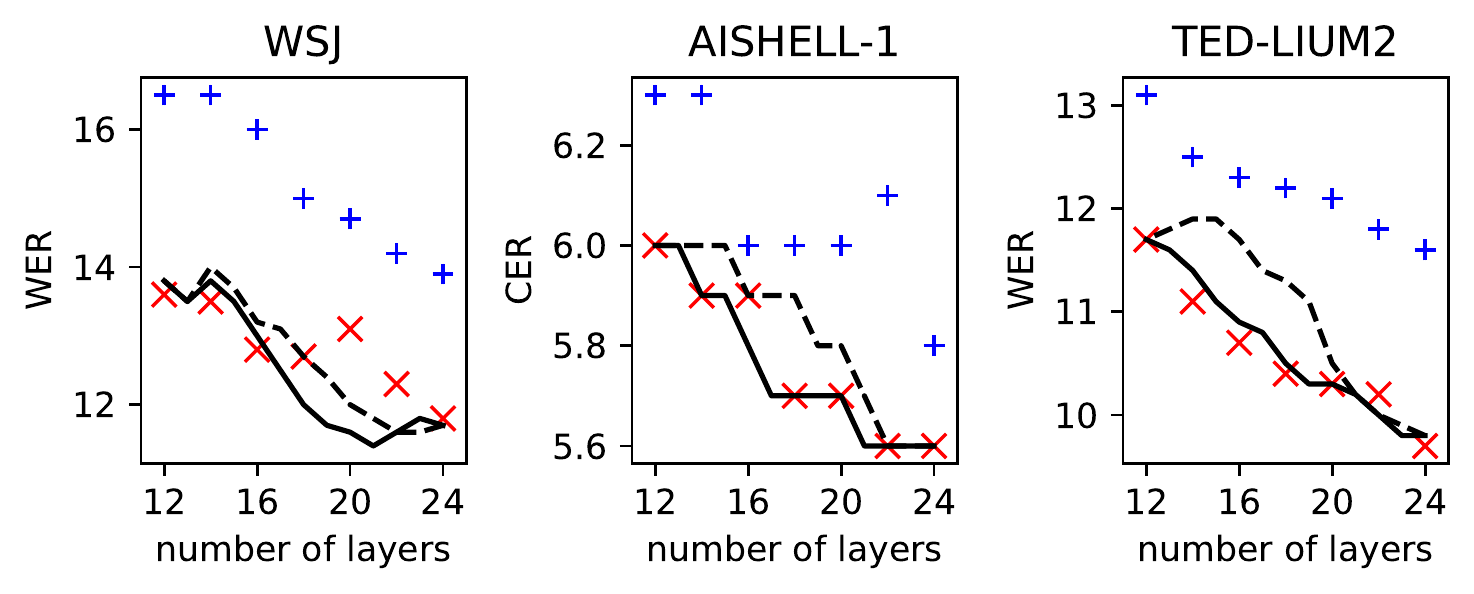}
  \vspace{-8mm}
\caption{Word error rates (WERs) for WSJ and TED-LIUM2 and character error rates (CERs) for AISHELL-1.
  {\color{blue}$+$}: baseline A, {\color{red}$\times$}: baseline B, dashed: intermediate sub-models, solid: sub-models found by iterative search.
  Note that all sub-model is induced from the one model without fine-tuning.
  See Section~\ref{sec:main_results} for details.
  }
  \label{fig:main_results}
  \vspace{-6mm}
\end{figure}

Figure~\ref{fig:main_results} shows the word error rates (WERs) for WSJ and TED-LIUM2 and character error rates (CERs) for AISHELL-1.
The baselines are displayed as individual marks and pruned sub-models are displayed as lines.
We first see the pruning-aware model and 24-layer baseline B have nearly same error rates.
This indicates that the proposed pruning method does not harm the full model performance.
Also, the half-sized sub-model and 12-layer baseline B also have nearly same error rates.
Iterative pruning reaches the individual baseline B models for most of cases.
This shows the effectiveness of the suggested method.

\begin{figure}[t]
  \centering
  \includegraphics[width=0.7\linewidth]{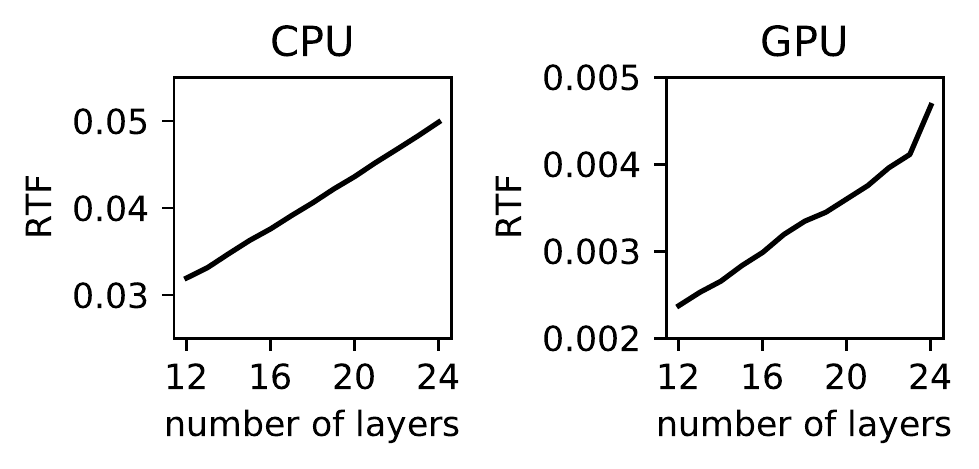}
  \vspace{-4mm}
\caption{Real-time factor (RTF) of the proposed model for varying depth. See Section~\ref{sec:main_results} for details.}
  \label{fig:rtf}
  \vspace{-4mm}
\end{figure}

We measure the real-time factor (RTF) of the pruning-aware model for each number of layers.
We use 1 Xeon CPU of 2.2GHz and 1 P40 GPU respectively.
Figure~\ref{fig:rtf} shows that the proposed model is very fast (RTF 0.05 on CPU and 0.005 on GPU),
due to lightweight and non-autoregressive properties of CTC,
and it can further improve RTF (2.5x on GPU) at the cost of small degradation on WER (18\%).

\subsection{Ablation study}
\label{sec:ablation_study}

\begin{figure}[t]
  \centering
  \includegraphics[width=0.7\linewidth]{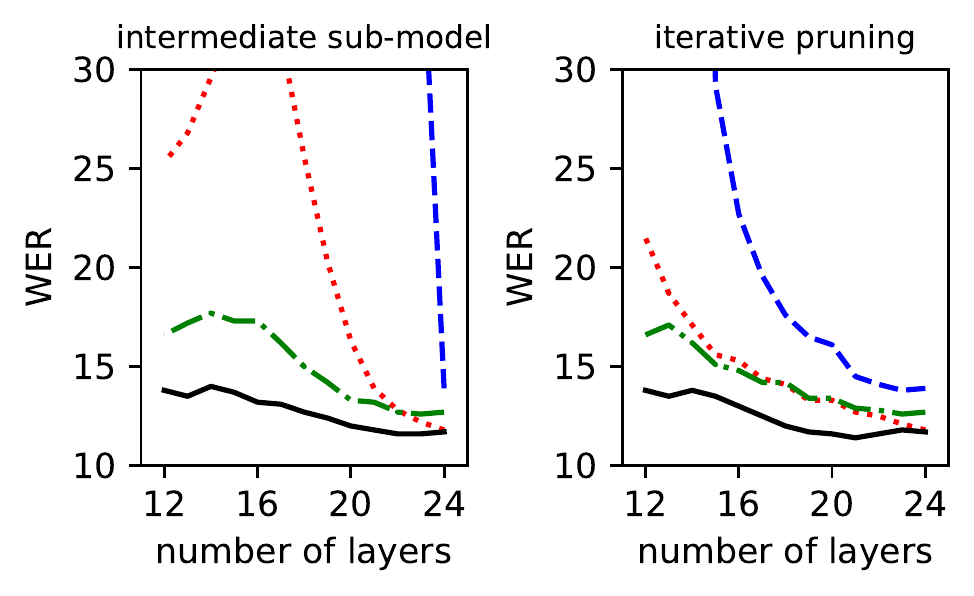}
  \vspace{-4mm}
\caption{Word error rates (WERs) for WSJ for sub-models induced from 24-layer model.
  Solid: proposed model,
  {\color{blue}dashed}: baseline A,
  {\color{red}dotted}: baseline B.
  {\color{ForestGreen}dashdot}: baseline B with in-place distillation.
  See Section~\ref{sec:ablation_study} for details.
  }
  \label{fig:ablation_study}
  \vspace{-4mm}
\end{figure}

In this section, we show how the change of intermediate CTC proposed at Section~\ref{sec:svcca_variants} improves sub-models.
We prune 24-layer baseline models of Section~\ref{sec:main_results} and compare their performance to the proposed pruning-aware model.
Figure~\ref{fig:ablation_study} shows the WER of sub-models induced by each model,
either by intermediate strategy (left) or iterative pruning (right).

We see that baseline A (dashed), without any regularization, does not work well with pruning.
It is much improved by baseline B (dotted), but it is still not close to individual baselines.
We also see that, for layers between the intermediate and final one, the intermediate sub-model performs poorly.

Finally, the proposed model (solid) gives huge improvement over baseline B, implying that the adjustment of intermediate CTC is very important for pruning, although it does not affect the performance of the full model.

As a preliminary experiment, we also apply in-place distillation~\cite{yu2019universally} to the baseline B. The result (dashdot) shows that, while it improves intermediate sub-models, it does not improve the iteratively pruned sub-models and degrades the full model.

\section{Conclusion}
\label{sec:conclusion}

We present a training method for CTC-based ASR models, which allows reducing model depth after training, without any fine-tuning.
The method is based on intermediate CTC and stochastic depth, previously used as regularization during training.
We investigate the behavior of the regularization methods using SVCCA and
improve intermediate CTC for layer pruning while retaining the regularization power.
We empirically show the 24-layer Transformer CTC model can be pruned to 12-layer which reaches the performance of models trained from scratch.

\bibliographystyle{IEEEtran}
\bibliography{mybib}

\end{document}